\documentclass[final,5p,times,twocolumn]{elsarticle}

\usepackage{amssymb}

\journal{Nuclear Physics B}

\begin{document}

\begin{frontmatter}

\title{Measurement of the charged pion mass using X-ray spectroscopy of exotic atoms}

\author[1]{M.\,Trassinelli\corref{cor1}}
\author[2]{D.\,F.\,Anagnostopoulos}
\author[3]{G.\,Borchert\fnref{pa1}}
\author[4]{A.\,Dax}
\author[5]{J.-P.\,Egger}
\author[3]{D.\,Gotta}
\author[3]{M.\,Hennebach\fnref{pa2}}
\author[6]{P.\,Indelicato}
\author[4]{Y.-W.\,Liu\fnref{pa3}}
\author[6]{B.\,Manil\fnref{pa4}}
\author[7]{N.\,Nelms\fnref{pa5}}
\author[4]{L.\,M.\,Simons}
\author[7]{A.\,Wells}

\cortext[cor1]{Corresponding author}
\fntext[pa1]{present address: TU Munich, D-85747 Garching, Germany}
\fntext[pa2]{present address: DAHER NUCLEAR TECHNOLOGIES GmbH, D-63457 Hanau, Germany}
\fntext[pa3]{present address: Phys. Depart., National Tsing Hua Univ., Hsinchu 300, Taiwan}
\fntext[pa4]{present address: Lab. de Physique des Lasers, Universit\'e Paris 13, Sorbonne Paris Cit\'e, CNRS, France}
\fntext[pa5]{present address: ESA-ESTEC, PO Box 299, 2200 AG, Noordwijk, The Netherlands}

\address[1]{Institut des NanoSciences de Paris, CNRS-UMR 7588, Sorbonne Universit\'es, UPMC Univ Paris 06, 75005, Paris, France}
\address[2]{Dept. of Materials Science and Engineering, University of Ioannina, GR-45110 Ioannina, Greece}
\address[3]{Institut f\"ur Kernphysik, Forschungszentrum J\"ulich GmbH, D-52425 J\"ulich, Germany}
\address[4]{Laboratory for Particle Physics, Paul Scherrer Institut, CH 5232-Villigen PSI, Switzerland}
\address[5]{Institut de Physique de l'Universit\'{e} de Neuch\^{a}tel, CH-2000 Neuch\^{a}tel, Switzerland}
\address[6]{Laboratoire Kastler Brossel,  Sorbonne Universit\'es, UPMC Univ. Paris 06, Case 74; 4, place Jussieu, 75005 Paris, France
and
Laboratoire Kastler Brossel,  CNRS, 75005, Paris, France
and
Laboratoire Kastler Brossel, D\'epartement de Physique de l'\'Ecole Normale Sup\'erieure,  24 Rue Lhomond, 75005, Paris, France}
\address[7]{Dept. of Physics and Astronomy, University of Leicester, Leicester LEI7RH, England}

\begin{abstract}
The $5g-4f$ transitions in pionic nitrogen and muonic oxygen were measured simultaneously by using a gaseous nitrogen-oxygen mixture at 1.4\,bar. Due to the precise knowledge of the muon mass the muonic line provides the energy calibration for the pionic transition. A value of (139.57077\,$\pm$\,0.00018)\,MeV/c$^{2}$ ($\pm$\,1.3ppm) is derived for the mass of the negatively charged pion, which is 4.2ppm larger than the present world average. 
\end{abstract}


\end{frontmatter}

X-ray spectroscopy of exotic atoms allows the determination of the mass of captured negatively charged particle like muons, pions, and antiprotons from the energies of the characteristic X-radiation. X-ray  transitions occur during the de-excitation cascade of the exotic atom which is formed at principal quantum numbers of $n\approx 16$ in the case of pions\,\cite{ERICE89,Got04}. The precise determination of the pion mass requires the use of X-ray lines which are not affected either by strong-interaction effects nor by collisions with surrounding atoms. Such conditions are found in the intermediate part of the cascade for exotic atoms formed in gases.

The most recent X-ray measurements were performed at the Paul Scherrer Institut (PSI) and used either a DuMond\,\cite{Jec86a,Jec86b,Jec94} or a Johann-type crystal spectrometer\,\cite{Len98}. In the case of the DuMond spectrometer, the energy calibration for the pionic magnesium $(4f-3d)$ transition was performed with a nuclear $\gamma$-ray, while for the Johann set-up K$\alpha$ fluorescence radiation from copper was used to determine the energy of the pionic nitrogen $(5g-4f)$ transition. 

In the $\pi$Mg experiment, electron refilling is unavoidable due to the use of a solid state target. Different assumptions on the K electron population lead to differences in the pion mass up to 16ppm\,\cite{Jec94}. 
The previous $\pi$N experiment, as well as the present one, used a nitrogen gas target at pressures around 1~bar, where electron refilling is unlikely\,\cite{Bac89,Kir99}, {\it i.\,e.} the de-excitation cascade is decoupled from the environment. The absence of refilling of the electrons ejected already during the upper part of the cascade by internal Auger effect manifests in the appearance of X-ray lines at $n\geq5$, which otherwise would be converted into Auger transitions\,\cite{Bur53,Vog80,Bac85}. Furthermore, a large Doppler broadening was measured for $(5-4)$ transitions\,\cite{Sie00}. It originates from Coulomb explosion during the formation process of the exotic atom  with molecules and indicates that the velocity at the time of X-ray emission is essentially unchanged since the breakup of the molecule. Thus, the absence of screening effects from remaining electrons in the intermediate part of the atomic cascade leads to a unique solution for the mass\,\cite{Len98}. In addition, in dilute targets the line intensity is already mostly collected in the circular transitions $(n,\ell =n-1)\rightarrow (n-1,\ell =n-2)$, where corrections owing to the hadronic potential are still tiny. 

From the $\pi$N experiment m$_{\pi^{-}}=(139.57071 \pm 0.00053)$\,MeV/c$^2$ is obtained which suggests that both K electrons are present when the $\pi$Mg$(4f-3d)$ transition occurs (solution B: m$_{\pi^{-}}=(139.56995 \pm 0.00035)$\,MeV/c$^2$). This is corroborated by the fact that the result, assuming 1 K electron only (solution A: m$_{\pi^{-}} = (139.56782 \pm 0.00037)$\,MeV/c$^2$), is in conflict with the measurement of the muon momentum for charged pion decay at rest $\pi^{+}\rightarrow \mu^{+}\nu_{\mu}$\,\cite{Ass96}. For solution A, the mass squared of the muon neutrino becomes negative by six standard deviations, whereas the average of solution B and the result of the $\pi$N$(5g-4f)$ measurement (m$_{\pi^{-}}=(139.57018 \pm 0.00035)$\,MeV/c$^2$\,\cite{Len98}) yields the upper limit m$_{\mu_{\nu }}<\,190$\,keV/$c^{2}$ (90\% c.\,l.)\,\cite{PDG14}. 
 
The experiment described here resumes the strategy of the gas target, but exploits (i) the high precision of 0.033ppm for the mass of the positively charged muon being m$_{\mu^{+}} = (105.6583715 \pm 0.0000035)$\,MeV/c$^2$\,\cite{PDG14} and (ii) the unique feature that in $\pi$N and $\mu$O transition energies almost coincide (Tab.\,\ref{table:energies}).  
Using a N$_2$/O$_2$ gas mixture in the target allows the simultaneous measurement of $\pi$N and $\mu$O lines, with the muonic transition serving as an on-line calibration. Hence, systematic shifts during the unavoidably long measuring periods are minimized.

In the case of nitrogen and oxygen, $(6h-5g)$, $(5g-4f)$, and $(4f-3d)$ transitions meet the operating conditions of the crystal spectrometer. Finally, the $(5g-4f)$ transition was chosen because: (i) for the   $(6h-5g)$ lines (2.2\,keV) absorption in the target gas itself and windows significantly reduces the count rate and (ii) the $3d$-level energy in $\pi$N requires a substantial correction because of the strong interaction. Electromagnetic transition energies (Tabs.\,\ref{table:energies} and \ref{table:energies_angles}) were calculated using a multi-configuration Dirac-Fock approach\,\cite{Des03,San05} to a precision of $\pm 1$\,meV and include relativistic and quantum electrodynamics contribution (relativistic recoil, self-energy, vacuum polarization) as well as the hyperfine structure of pionic nitrogen\,\cite{Tra07}.

Energy shifts due to nuclear finite size are found to be as small as 4\,aeV and 2\,peV for the $5g$ and $4f$ levels in $\pi$N. Values for nuclear masses, radii, and moments were taken from recent compilations\,\cite{Aud03,Ang04,Rag89}. The strong-interaction shifts of the $\pi$N  levels were estimated from interpolating the measured hadronic $2p$-level shifts in $\pi$C and $\pi$O\,\cite{Cha85} and by using scaling relations based on the overlap of nucleus and a hydrogen-like wave function for the pion orbit (see Tab.\,\ref{table:errors}).  Details on the  calculation of the transition energies may be found elsewhere\,\cite{Tra05}.

\setlength{\tabcolsep}{7mm}
\begin{table*} 
\begin{center} 
\caption{Calculated contributions to the total QED transition energy of $\mu$O and $\pi$N $(5g-4f)$ lines (in eV)\,\cite{Tra07}. For the pionic transition, the world average pion mass value as given in\,\cite{PDG14} is used. The $\mu$O line constitutes a triplet due to the muon spin. The total uncertainty of the QED calculation (excluding the uncertainty of the pion mass) is $\pm$1\,meV.} 
\label{table:energies} 
\begin{tabular}{lrrrr} 
\\[-3mm]\hline\\[-3.0mm]
               & \multicolumn{3}{c}{$\mu^{16}$O}                                     & \multicolumn{1}{c}{$\pi^{14}$N}\\ 
transition     & ($5g_{9/2}-4f_{7/2}$) & ($5g_{7/2}-4f_{7/2}$) & ($5g_{7/2}-4f_{5/2}$) & ($5g-4f$) \\
\hline\\[-3mm]
Coulomb                        &  4022.8625&  4022.6188&  4023.4124~~~&  4054.1180 \\ 
self energy                    & --\,0.0028& --\,0.0013& --\,0.0013~~~& --\,0.0001 \\ 
vac.\,pol.\,(Uehling)          &     0.8800&     0.8800&     0.8807~~~&     1.2485 \\ 
vac.\,pol.\,Wichman-Kroll      & --\,0.0007& --\,0.0007& --\,0.0007~~~& --\,0.0007 \\ 
vac.\,pol.\,two-loop\,Uehling  &     0.0003&     0.0004&     0.0004~~~&     0.0008 \\ 
vac.\,pol.\,K\"all\'en-Sabry   &     0.0084&     0.0084&     0.0084~~~&     0.0116 \\ 
relativistic\,recoil           &     0.0025&     0.0025&     0.0025~~~&     0.0028 \\ 
hyperfine\,structure           &    --~~~~~&    --~~~~~&    --~~~~~~~~& --\,0.0008 \\ 
\hline\\[-4mm]
Total                          & 4023.7502 &  4023.5079&  4024.2983~~~&  4055.3801    \\ 
\hline\\[-7mm]
\end{tabular} 
\end{center} 
\end{table*}

\begin{figure}[th!]
\begin{centering}
\includegraphics[scale=1.18]{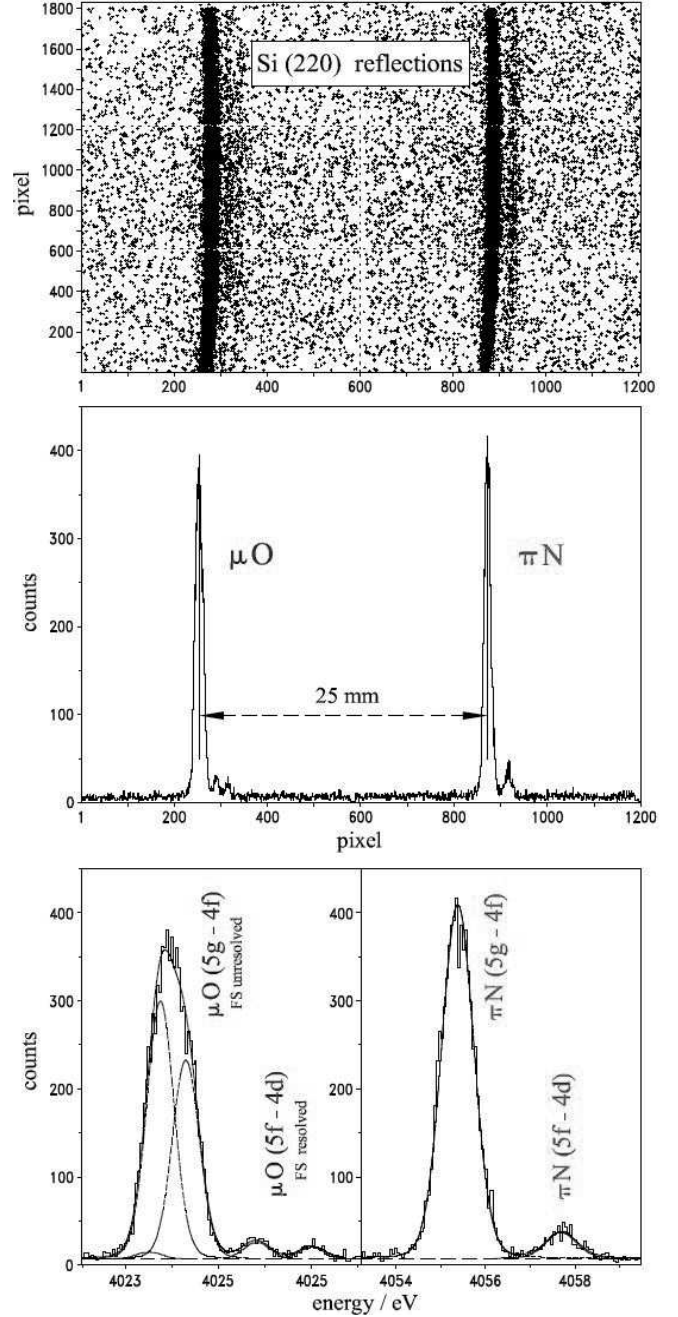}
\caption{Simultaneously measured $(5g-4f)$ transitions in muonic oxygen (calibration) and pionic nitrogen. Top: Distribution of the Bragg reflections on the surface of the $2\times 3$ CCD array. The binning corresponds to the pixel size of the CCDs (note the different scales vertically and horizontally). Straight dashed lines indicate CCD boundaries. Middle: Projection on the axis of dispersion after correction for curvature (see text). Bottom: Details of the fit to line patterns.}
\label{fig:muO_piN}
\end{centering}
\end{figure}

The measurement was performed at the high-intensity pion beam line $\pi$E5 of the Paul Scherrer Institute (PSI) using a set-up similar to the one used by Lenz et al.\,\cite{Len98}. Major improvements are: (i) The use of cyclotron trap II\,\cite{Sim88} having a larger gap between the magnet coils yielding a substantially increased muon stop rate, (ii)  a Bragg crystal of superior quality and (iii) a large-area X-ray detector in order to simultaneously cover the reflections of the muonic and pionic transitions  (see Fig.\,\ref{fig:muO_piN}). In addition, the average proton current of the accelerator was about 1.4\,mA, which is 40\% higher than in the previous experiment. 

The N$_2$/O$_2$ gas mixture was enclosed in a cylindrical target cell placed at the centre of the cyclotron trap. The cell wall was made of a 50\,$\mathrm{\mu}$m thick Kapton$^{\small{\textregistered}}$ foil. Towards the crystal spectrometer a circular 7.5\,$\mathrm{\mu}$m Mylar$^{\small{\textregistered}}$ window was used supported by a stainless steel honeycomb structure with a free area of 90\%. The target was operated at 1.4\,bar and room temperature.

The muons used originate from the decay of slow pions inside the cyclotron trap, because the stop density for muons at the high-intensity pion beam is still superior to the one at a dedicated muon channel. For the simultaneous measurement comparable count rates are required for the $\pi$N and the $\mu$O line. This was achieved with a N$_{2}$/O$_{2}$ mixture of 10\%/90\% by adapting the set of polyethylene degraders inside the magnet gap and optimized by means of an X-ray  measurement using a Si(Li) semiconductor detector. 

The crystal spectrometer is set up in Johann geometry\,\cite{Joh31} using a spherically bent Bragg crystal and optimised to the needs of exotic-atom X-ray spectroscopy\,\cite{Got16}. Such a configuration allows the simultaneous measurement of two different energies within an energy interval, the limits of which are given by the extension of the target in the direction of dispersion and correspondingly by the size the detector. Spherical bending leads to a partial vertical focussing\,\cite{Egg65} which increases the count rate. 

The Bragg crystal was made from a silicon crystal disk of 290\,$\mathrm{\mu}$m thickness and of a diameter of 100\,mm. The disk is attached to a high-quality polished glass lens defining a spherical segment. The average radius of curvature of the crystal surface was measured to $R_{\mathrm{c}}=(2981.31\,\pm\,0.33)$\,mm by sampling 500 points at the surface with a mechanical precision sensor (performed by Carl Zeiss AG, D-73447 Oberkochen, Germany). An upper limit for the cut angle (angle between crystal surface and reflecting lattice planes) was determined in a dedicated measurement to be 120 seconds of arc\,\cite{Cov08}. Hence, the focal condition corresponds to the symmetric Bragg case being $R_{\mathrm{c}}\cdot \sin\Theta_{\mathrm{B}}$. The measurement uses the second order reflection at the (110) planes. An aluminium aperture of 90\,mm diameter covered the boundary region of the Si disk in order to avoid edge effects. For source geometry as given here, the overall efficiency of the crystal set-up is $\approx5\cdot 10^{-8}$. About 85\% of the reflected intensity is covered by the sensitive area of the detector.

The detector with a total sensitive area of about 48$\times$72\,mm$^{2}$ (width$\times$height) was built up by a 2$\times$3 array of charge-coupled devices (CCDs) of 24\,mm$\times$24\,mm (600$\times$600 pixels) with frame storage option\,\cite{Nel02}. Having a depletion depth of about 30\,$\mathrm{\mu m}$ these CCDs reach their maximum in detection efficiency of almost 90\% at 4\,keV. The detector surface is oriented perpendicular to the direction of the incoming X-rays. Excellent background conditions are achieved (i) by using an especially tailored  concrete shielding of at least 1\,m thickness between the X-ray detector and the target region and (ii) by exploiting the different pixel topology of low-energy X-rays and background events, which are mainly caused by neutron induced high energetic nuclear $\gamma$ rays \,\cite{Got04,Len98}.

The Bragg angle for the $\pi$N$(5g-4f)$ transition and thereby its energy is determined from the position difference to the $\mu$O$(5g-4f)$ line. The positions are determined from the projection of the pattern on the CCD onto the direction of dispersion after correction for curvature by means of a parabola fit (Fig.\,\ref{fig:muO_piN}). The main transitions $\mu$O$(5g-4f)$ and $\pi$N$(5g-4f)$ are separated by about 25\,mm. 

About 9000 events for each element were collected in each of the $(5g-4f)$ transitions during 5~weeks of data taking.  The count rates for the $\pi$N and $\mu$O transitions were about 15 events per hour each. Only a common small drift was observed for the line positions of less than one pixel in total. Because of the simultaneous measurement the position difference is not affected. Bragg angle dependent corrections are small because the leading order cancels in such a difference measurement performed in the same order of reflection. 

\setlength{\tabcolsep}{2.2mm}

\begin{table}[h]
\begin{center}
\caption{Transition energies $E_{\mathrm {QED}}$\,\cite{Tra07}  and Bragg angles $\Theta_{\mathrm{B}}$ of the $\mu$O and $\pi$N lines used in the fit to the spectrum. The relative intensities within the fine structure multiplets of $\mu$O (FS int.) have been fixed in the fit to the statistical weight. The Bragg angle includes the index of refraction shift calculated with the code XOP\,\cite{San98}. For twice the lattice distance $2d=0.768\,062\,286\,(13)$\,nm is assumed at a temperature of 22.5$^{\circ}$C\,\cite{CODATA}. The conversion constant used is $hc=1.239\,841\,930\,(28)\,$nm$\cdot$keV\,\cite{PDG14}. The $\pi$N$(5g-4f)$ and $\pi$N$(5f-4d)$ transition energies include the strong-interaction shift (see tab.\,\ref{table:errors}).}
\label{table:energies_angles}
\begin{tabular}{lccc}
\hline\\[-3.5mm]
transition                      & FS int.&   $E_{\mathrm {QED}}$/eV&   $\Theta_{\mathrm{B}}$  \\           
\hline\\[-3mm]
$\mu^{16}$O$(5g_{7/2}-4f_{7/2})$ & 1& 4023.5079 &  53$^{\circ}$21'51.48"  \\
$\mu^{16}$O$(5g_{9/2}-4f_{7/2})$ &35& 4023.7503 &  53$^{\circ}$21'34.77"  \\
$\mu^{16}$O$(5g_{7/2}-4f_{5/2})$ &27& 4024.2984 &  53$^{\circ}$20'57.01"  \\[1mm]

$\mu^{16}$O$(5f_{5/2}-4d_{5/2})$ & 1& 4025.3956 &  53$^{\circ}$19'41.47"  \\
$\mu^{16}$O$(5f_{7/2}-4d_{5/2})$ &20& 4025.8031 &  53$^{\circ}$19'13.44"  \\
$\mu^{16}$O$(5f_{5/2}-4d_{3/2})$ &14& 4026.9922 &  53$^{\circ}$17'51.70"  \\[1mm]

$\mu^{16}$O$(5d_{5/2}-4p_{3/2})$ & 9& 4028.5625 &  53$^{\circ}$16'3.90"  \\
$\mu^{16}$O$(5d_{3/2}-4p_{1/2})$ & 5& 4033.5273 &  53$^{\circ}$10'24.10"  \\[1mm]

$\mu^{18}$O$(5g_{7/2}-4f_{7/2})$ & 1& 4026.6692 &  53$^{\circ}$18'13.90"  \\
$\mu^{18}$O$(5g_{9/2}-4f_{7/2})$ &35& 4026.9132 &  53$^{\circ}$17'57.13"  \\
$\mu^{18}$O$(5g_{7/2}-4f_{5/2})$ &27& 4027.4642 &  53$^{\circ}$17'19.28"  \\
\hline\\[-3mm]
$\pi^{14}$N$(5g-4f)$             &  & 4055.3802 &  52$^{\circ}$45'46.76" \\
$\pi^{14}$N$(5f-4d)$             &  & 4057.6984 &  52$^{\circ}$43'11.81" \\
\multicolumn{2}{l}{$\pi^{14}$N$(5d-4p)$ ~~~{\it QED only}}\             &   4061.9460 &  52$^{\circ}$38'28.76" \\[1mm]
$\pi^{15}$N$(5g-4f)$             &  & 4058.2394 &  52$^{\circ}$42'35.67" \\
$\pi^{15}$N$(5f-4d)$             &  & 4060.5605 &  52$^{\circ}$40'~0.95" \\
\hline\\[-2mm]

\end{tabular}\\
\end{center}
\end{table}

In fourth order, the Bragg angles of the Cu K$\alpha$ lines are very close to the ones of the $\mu$O$(5g-4f)$ transitions. Therefore, in addition Cu X-rays were repeatedly recorded as a stability monitor corroborating the amount of the small common drift observed for the $\mu$O/$\pi$N pair. 

Various parameters of the analysis and of the set-up enter in the determination of the line positions and their difference. These contributions and their uncertainties are summarised in Table\,\ref{table:errors} and are discussed in detail below.

\setlength{\tabcolsep}{7.6mm}

\begin{table*}[t]
\begin{center}
\caption{Corrections to the measured angle difference between the $\pi^{14}$N$(5g-4f)$ and the $\mu^{16}$O$(5g_{9/2}-4f_{7/2})$ transitions and associated uncertainties. A 1 ppm change in the pion mass corresponds to 4.055\,meV in transition energy, to 0.27 arcsec in diffraction angle, or to a displacement of 3.2\,$\mu$m in the detector plane. Contributions to the mass uncertainty from lattice and conversion constant cancel in leading order because the measurement principle is based on the angular difference. For more details see text.}
\label{table:errors}
\begin{tabular}{lcccc}
\\[-6.5mm]
\hline\\[-3.5mm]
 type of uncertainty            & $\mu$O  & $\pi$N  & total     & uncertainty \\           
                                & / arcsec& / arcsec&  / arcsec &  / ppb      \\
\hline\\[-3mm]
index of refraction shift       & 13.22   & 12.94   & -\,0.28   & $\pm\,20$ \\
silicon lattice constant        &         &         &           &    $\pm\,2$        \\
bending correction              & 14.01   & 13.71   & 0.30      & $\pm\,20$ \\
penetration depth correction    & -0.07   & -0.07   & 0         & $\pm\,4$ \\[1mm]
focal length                    &         &         &           & $\pm\,670$ \\
CCD alignment                   &         &         &           & $\pm\,340$ \\
pixel distance                  &         &         &           & $\pm\,120$ \\[1mm]
alignment of detector normal    &         &         &           & ${+~\,0\atop -\,30}$ \\[1mm] 
detector height offset          &         &         &           & ${+~\,0\atop -\,35}$ \\[1mm]
shape of target window          &         &         &           & $\pm\,100$ \\
shape of reflection             &         &         &           & $\pm\,225$ \\
individual curvature correction &         &         &           & $\pm\,150$ \\
temperature correction          &         &         &           & $\pm\,30$ \\[1mm]
response function and Doppler broadening  &   &     &           & ${+\,290\atop -\,350}$ \\[1mm] 
line pattern modelling              &   &       &        & ${+\,190\atop -\,290}$ \\
fit interval                        &   &       &        & $\pm\,15$ \\
\hline\\[-3mm]
QED energy                          &   &       &        & $\pm\,350$ \\                
conversion constant $hc$            &   &       &        &    $\pm\,2$        \\
$4f$ strong interaction 45\,$\mu$eV &   & 0.003 & -0.003 & $\pm\,10$ \\
$5g$ strong interaction 0.2\,$\mu$eV&   & 0.000 &  0.000 & $\pm\,0$ \\
K electron screening                &   &       &        & $\pm\,0$ \\[1mm] 
\hline\\[-2.5mm]
total systematic error              &   &       &        & ${+~\,950\atop -\,1000}$ \\[1mm] 
statistical error                   &   &       &        & $\pm\,820$ \\ 
\hline\\[-4mm]
\end{tabular}\\
\end{center}
\end{table*}

\paragraph{\bf Index of refraction shift} 
The systematic uncertainty of the index shift correction is assumed to be about 5\%\,\cite{Hen93,Cha95}, {\it i.\,e.} the uncertainty of the difference is negligibly small.

\paragraph{\bf Silicon lattice constant and wavelength conversion} 
Both the silicon lattice constant $2d$ and the conversion constant $hc$ are known to an accuracy of $\approx\,10^{-8}$.

\paragraph{\bf Bending and penetration depth corrections} 
The energy dependent penetration depths of the X-rays lead to different corrections for the lattice constant of the Bragg crystal due to its curvature. The difference of the shift due to the average penetration depths itself turns out to be negligible. The primary extinction lengths including absorption were calculated both with the codes XOP\,\cite{San98} and DIXI\,\cite{Hoe98}, where results were found to coincide perfectly. We assume that the crystal behaves like an ideal one for such large bending radii\,\cite{Usc93}. The corrections for the Bragg angle were calculated following the approach of\,\cite{Cem92,Cha95a} using for the Poisson number the value $\nu = 0.208$ obtained from\,\cite{Wor65,Chu96}. 

\paragraph{\bf Focal length} 
Because of the different focal lengths for the $\pi$N and $\mu$O lines of 18.4\,mm, the detector was placed in an intermediate position, which was determined by a survey measurement to be (2388.27\,$\pm$\,0.20)\,mm. The uncertainty of the distance crystal-to-detector represents the largest contribution to the systematic error.

\paragraph{\bf CCD alignment and pixel distance} 
In the CCD array small gaps of the order of 0.3\,mm emerge between the individual devices. Secondly, the nominal pixel size of the CCDs, reported to be $40\,\mathrm{\mu m}\times\,40\,\mathrm{\mu m}$ at room temperature, changes for the operating temperature of $-\,100^{\circ}$C. Both the relative orientations of the six CCD devices and the average pixel distance have been measured precisely in a separate experiment using a nanometric quartz mask\,\cite{Ind06}. The average pixel distance was found to be $(39.9775\,\pm\,0.0006)\,\mathrm{\mu m}$, substantially different from the nominal value.

\paragraph{\bf Alignment of detector normal} 
The surface of the CCD array was set-up perpendicular to the direction crystal-detector to better than $\pm\,0.14^{\circ}$. The uncertainty also includes the imperfectness of the vacuum tubes, of their connections,  and of the support structures of the CCDs.

\paragraph{\bf Detector height offset} 
A possible offset in height of the detector from the ideal geometry defined by the plane through the centres of X-ray source, crystal, and detector leads to a distortion of the reflections. The size of such an effect was quantified by means of a Monte-Carlo simulation. 

\paragraph{\bf Shape of the target window} 
The circular shape of the target window leads to boundaries of different inclination for the $\pi$N and $\mu$O reflections. The corresponding possible uncertainty for the position difference was determined from a Monte-Carlo simulation.

\paragraph{\bf Shape of reflection}
The curvature of the $\pi$N and $\mu$O reflections is determined from a parabola fit to the hit pattern of the circular transitions. The assumption of a parabolic shape for the curvature is valid only close to the above-mentioned central plane. In addition, the curvature fit assumes a constant width of the reflection. A possible effect on the position difference over the height of the CCD array, which principally increases with increasing distance from the central plane, was studied by restricting the detector surface in height. The deviations are found to be far below the statistical error of the line positions.

\paragraph{\bf Individual curvature correction} 
The parabola parameters for the $\pi$N and $\mu$O reflections are slightly different because of different focal lengths. No difference could be verified from the fits which, however, is expected within the available statistics. The uncertainty is therefore given by the error of the fit to the curvature. For curvature correction, the average values were taken of  the $\pi$N and $\mu$O reflection. 

\paragraph{\bf Temperature correction} 
The temperature during the measurement varied between 19$^{\circ}$C and 21$^{\circ}$C during the measurement. All periods were rescaled to 22.5$^{\circ}$C by using the appropriate thermal expansion coefficient. The main correction comes from the change of the lattice constant. A smaller contribution arises from the variation of the distance crystal detector.

\paragraph{\bf Response function and Doppler broadening} 
The response is found by a convolution of the intrinsic crystal response with the aberration caused by the imaging properties of a spherically bent crystal. The crystal response was calculated with the code XOP\,\cite{San98}, and the geometry was taken into account by means of Monte-Carlo ray-tracing\,\cite{Got16}. The resulting response shows a significant asymmetry having a width of 450\,meV (FWHM). 

Measured line widths of $\pi$N and $\mu$O transitions, however, are dominated by Doppler broadening due to Coulomb explosion\,\cite{Sie00}, which was underestimated in the analysis reported by Lenz et al.\,\cite{Len98} because of an inferior quality of the Bragg crystal. The line shapes are almost symmetric having a width of about 750\,meV (FWHM). The Doppler broadening was accounted for best by folding in an additional Gaussian of about 40 seconds of arc. The Gaussian was  determined from the analysis of a dedicated measurement optimised for pion stops, where in total 60000 events were accumulated in the $\pi$N$(5g-4f)$ transition. 

The defocusing due to the different focal lengths is included in the Monte-Carlo based response, which is calculated for the appropriate distance in each case. In addition, it was verified that the parameters found in the curvature fit to the data are reproduced for the Monte-Carlo result.

\paragraph{\bf Line pattern} 
The total line pattern to be considered is a superposition of the circular $(5g-4f)$ and the inner transitions $(5f-4d)$ and  $(5d-4p)$ together with the corresponding contributions from the other isotopes (Tab.\,\ref{table:energies_angles}). The isotope abundances are fixed as tabulated ($^{16}$O/$^{18}$O: 99.76\%/0.21\%, $^{14}$N/$^{15}$N: 99.64\%/0.36\%). The relative intensities of the inner transitions are due to the cascade dynamics and, therefore, free parameters of the fit.

The line positions within the $\pi$N and $\mu$O $(5g-4f)$ patterns were fixed according to the QED energies. In the case of $\mu$O, all fine structure components were included in the fit. For a proper description of the background, the two strong components of the $\mu^{16}$O$(5d-4p)$ triplet and the $\pi$N$(5d-4p)$ transition were included in the fit. For the pionic line, position and width were free parameters, because it is  shifted and broadened by about $1$\,eV compared to the electromagnetic value by the strong interaction\,\cite{Len98}.

\paragraph{\bf Fit interval} 
Changing the interval used in the fit of the line positions does affect the result insignificantly.

\paragraph{\bf K electron screening} 
From the analysis of the high-statistics $\pi$N$(5g-4f)$ data, we exclude the influence of satellites lines due to remaining K electrons. The energy shift of the pionic transition is calculated to be $-456$ ($-814$)\,meV in the case of one (two) K electron(s). Two hypothesis (presence of satellites or not) are compared via the Bayes factor \cite{Jay03,Siv06,Kas95,Gor07} yielding an upper limit of less than $3\cdot 10^{-6}$ for the relative intensity of possible satellites.

\hfill

The measured energy of the $\pi$N$(5g-4f)$ transition was found to be $(4055.3970\,\pm\,0.0033_{stat}\,\pm\,0.0038_{sys})$\,eV. Basically two facts limit the accuracy of the method of a simultaneous measurement as described here: (i) The low rate obtainable from the muonic transitions hinders to accumulate as high statistics as would be achievable when using a set-up optimised for pionic atoms. For pionic transitions, count rates being  a factor of 20 larger than for muonic X-rays can be achieved. (ii) The large Doppler broadening induced by  Coulomb explosion when using diatomic gases, which approximately doubles  the line width as expected from the spectrometer response. 

\begin{figure}[t]
\begin{center}
\includegraphics[scale=0.4,clip= true, trim = 0 120 20 120]{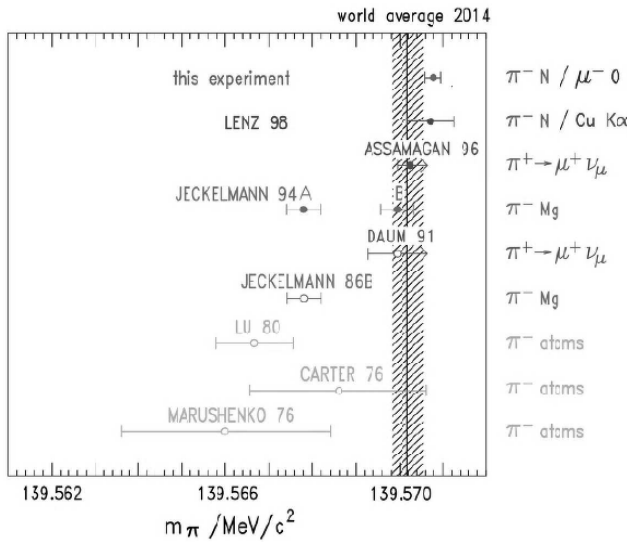}
\caption{Results for the mass of the charged pion. Also shown are previous exotic-atom results (Jeckelmann et al. (86B\,\cite{Jec86a,Jec86b}), Lu et al.\,\cite{Lu80}, Carter et al.\,\cite{Car76}, Marushenko et al.\,\cite{Mar76}) and $\pi^+$  decay at rest (Daum et al.\,\cite{Dau91}). The shaded region indicates the world average before this experiment\,\cite{PDG14}.}
\label{fig:pimass} 
\end{center}
\vspace{-6mm}
\end{figure}

To summarize, the mass of the negatively charged pion has been measured by means of equivalent X-ray transitions in hydrogen-like pionic nitrogen and muonic oxygen, where the muonic line serves as energy calibration. The value of (139.57077$\pm$\,0.00018)\,MeV/c$^{2}$ is 4.2ppm larger than the present world average\,\cite{PDG14}. Repeating the procedure as described in ref.\,\cite{Len98} by using the Cu K$\alpha_1$ line for calibration, yields a value of $m_{\pi}\,=\,$(139.57090\,$\pm$\,0.00056)~MeV/c$^{2}$. The accuracy of $\pm\,4.0$ppm represents the limit for a calibration with broad X-ray fluorescence lines. Both results are in good agreement with the mass obtained by\,\cite{Len98}, but 5.4ppm and 6.8ppm, respectively, above the result of the pionic magnesium experiment (solution B\,\cite{Jec94}) using a nuclear $\gamma$ ray for calibration (Fig.\,\ref{fig:pimass}). 

The analysis shows no evidence for any satellite lines from remaining electrons at the time of X-ray emission of the $(5g-4f)$ transition. This corroborates strongly our assumption for a complete depletion of the electron shell during the preceding steps of the atomic cascade.

In conclusion, the present study demonstrates the potential of  crystal spectroscopy with bent crystals in the field of exotic atoms. Its limits are given, on one hand, by statistics for the present beam and detector technologies. On the other hand, the systematic uncertainties discussed at length above  illustrate the level of sophistication which must be applied. 

Facing the fact that pion beams at PSI provide a flux of about $10^{9}$/s, the use of double-flat crystal  spectrometers may be considered allowing for absolute angle calibrations without a (muonic or X-ray) reference line. Choosing pionic transitions not affected by Coulomb explosion, {\it e.\,g.} from pionic neon, a precision for the pion mass determination of the order of 0.5ppm is feasible which, however, may be regarded to be the maximum achievable by means of exotic-atom X-ray spectroscopy. 

As a result, X-rays of hydrogen-like pionic atoms are useful to provide calibration standards in the few keV range, where suitable radioactive sources are not available\,\cite{Tra05,Ana03}. At present, the accuracy is given by the uncertainty of the pion mass\,\cite{Sch11}. The quality of such standards may benefit substantially from laser spectroscopy of metastable high-lying pionic states which is proposed to be performed in pionic helium also at PSI\,\cite{Hor14}.\\

Combined with the measurement of the muon momentum after pion decay at rest\,\cite{Ass96}, a non-zero value for the muon neutrino mass is obtained of $m_{\nu_{\mu}}=183\,{+\,62\atop -\,83}$ \,keV/c$^{2}$ (c.l. 90\%) when using the statistical approach of\,\cite{Fel98}. The result is far above the cosmological limit of at least 11\,eV/c$^{2}$ for the sum of all neutrino flavours\,\cite{PDG14}. However, extending the error limits to 3$\sigma$ either for the pion mass or the muon momentum yields values for m$_{\nu_{\mu}}$ consistent with zero. 

\hfill

We are grateful to N.\,Dolfus, H.\,Labus, B.\,Leoni and K.-P.\,Wieder for solving numerous technical problems. We thank the PSI staff for providing excellent beam conditions and appreciate the support by the Carl Zeiss AG, Oberkochen, Germany, which fabricated the Bragg crystals. We thank Prof. Dr. E.\,F\"{o}rster and his collaborators at the University of Jena, and A.\,Freund and his group at ESRF, for the help in characterising the crystal material as well as A. Blechmann for a careful study of the CCD performance. We are indebted to PSI for supporting the stay during the run periods (D.\,F.\,A.). This work is part of the PhD thesis of B.\,M. (Universit\'{e} Pierre et Marie Curie, 2001), N.\,N. (University of Leicester, 2002) and M.\,T. (Universit\'{e} Pierre et Marie Curie, 2005).

\end{document}